\documentclass[a4paper,11pt]{article}
\pdfoutput=1 % if your are submitting a pdflatex (i.e. if you have
\usepackage{jinstpub} % for details on the use of the package, please
\title{\boldmath Response to cosmic muons of scintillator-SiPM assemblies measured at different temperatures}

\author[a,1]{M.~Chadeeva,\note{Corresponding author.}}
\author[b]{ A.~Stifutkin}

\affiliation[a]{P.N. Lebedev Physical Institute of the RAS,\\53, Leninskyi prospekt, Moscow, Russia}
\affiliation[b]{NRNU MEPhI, \\31, Kashirskoe shosse, Moscow, Russia}

\emailAdd{chadeevamv@lebedev.ru}

\abstract{Highly segmented calorimeters represent a modern trend in experimental particle physics aimed at improving the energy resolution with the particle flow reconstruction. The widely used and cost-effective solution is the structure assembled from small scintillator elements readout by tiny photosensors. The improvement of signal to noise ratio can be achieved by operating at low temperatures. The paper presents studies of temperature dependence of response to muons for elements comprised of plastic scintillator tiles with different options of reflective coverage and read out by silicon photomultipliers. No temperature dependence of response for bare tiles has been detected. With reflective film, the reduction of response with decreasing temperature is observed. The measured effect amounts to $\sim$10\% in the range from $+$30\textdegree C to $-$30\textdegree C. }

\begin{document}
\maketitle
\flushbottom

\section{Introduction}
\label{sec:intro}
The highly granular calorimeters represent a sustainable trend in modern particle detector developments opening a possibility to implement the Particle Flow (PF) approach and improve jet energy resolution \cite{ref:pfa_brient,ref:pfa_morgunov,ref:pfa_thomson}. Many detector proposals for future experiments, e.g. CEPC \cite{ref:cepc} and FCC \cite{ref:fcc}, include highly granular calorimeter options and rely on the PF reconstruction in their performance studies. During the last two decades, the CALICE collaboration has realised a successful R\&D program for design, construction and tests of highly segmented electromagnetic and hadronic calorimeters based on different technologies and material combinations \cite{ref:Sefkow:2016}. An example of experimentally tested solution is the 1~m$^3$ CALICE AHCAL -- a sampling calorimeter assembled from 22 thousand 3M-wrapped plastic scintillator tiles as active elements readout by silicon photomultipliers (SiPM) \cite{ref:calice}. The combined silicon-scintillator technology has been also chosen for the CMS endcap calorimeter upgrade for the High-Luminosity LHC and the recently developed High Granularity CALorimeter (HGCAL) is now under construction \cite{ref:hgcal}. 

The usage of silicon photomultipliers in highly granular systems brings many advantages compared to vacuum devices -- low operating voltage and therefore low power consumption, small size allowing incorporation inside active elements without additional light guides and relatively high quantum efficiency. A typical detector element for high granularity option is a scintillator plate (tile) wrapped in reflective film and having a dimple for light to be directly readout by a SiPM. The wrapping helps to increase the light collection efficiency. However, the remaining problem is the SiPM noise, which is the most important factor affecting the performance especially in hard irradiation conditions~\cite{ref:garutti2019}. Operation at low temperatures helps to solve this problem, resulting typically in one order of magnitude dark count rate reduction per 40\textdegree C at the same overvoltage. For instance, the aforementioned HGCAL will be operated at $-$35\textdegree C. 

Different tests and calibrations of detector modules are done with minimum ionising particles including cosmic muons. For reliable projections of the performance estimates from the tests and calibrations at room temperature to operating conditions, the information about temperature dependencies of detector components would be very helpful. 
The temperature dependencies of breakdown voltages and other parameters of silicon photomultipliers are typically provided by the manufacturers. There are few published results describing the behaviour of several scintillator materials in different temperature ranges, where dependencies are observed to be quite weak \cite{ref:peralta,ref:carturan}. As for reflective films, the manufacturers state the independence of film strength and reflection quality of temperature in the range from $-$60\textdegree C up to $+$60\textdegree C. 
%The previous study of tiles wrapped in the 3M ESR film (3-mm thick trapezoid tiles for the HGCAL prototype) has shown the decrease of light yield with decreasing temperature by about 1 photoelectron per 10 degrees~\cite{ref:bull2025}. 
In this work, the temperature dependence of response to cosmic muons has been experimentally studied in the range from $-$30\textdegree C up to $+$30\textdegree C for scintillator-SiPM assemblies with bare scintillator tiles, scintillator tiles covered by reflective film slices and scintillator tiles wrapped in reflective film. 

\section{Materials and methods}
\label{sec:method}
The main objects of this research are small tiles produced from plastic scintillator. Two types of objects have been studied: the 3-mm thick square tiles of equal size 30$\times$30$\times$3~mm$^3$ (hereinafter called "AHCAL tiles") and
the 20-mm thick square tiles of equal size 38$\times$38$\times$20~mm$^3$ (hereinafter called "TEST tiles"). All tiles have dimples, 6~mm in diameter and 1.6~mm in depth, in the center of one big surface to accomodate SiPM and improve uniformity of light collection~\cite{ref:soldner}. The AHCAL tiles have been produced for the technological prototype  of the CALICE Analog Hadron Calorimeter at the Uniplast factory (Vladimir, Russia) by injection moulding technique from polystyrene with 1.5\% of PTP and 0.1\% POPOP dopants. They are machine wrapped in the 65-micron 3M Enhanced Specular Reflector Film (ESR). The TEST tiles have been made from scintillator blocks manufactured at the ISMA laboratory (Kharkiv, Ukraine) by extrusion technique from polystyrene with 2\% of PTP and 0.025\% POPOP dopants. The 20-mm thick perimeter sides of each TEST tile was manually coated by diffuse reflector (foamed polystyrene), these tiles have been measured with and without slices of 82-micron 3M ESR film on top and bottom of the big surfaces in attempt to disentangle the effects induced by scintillator and reflective film properties.

\subsection{Experimental setup}
\label{sec:setup}
\begin{figure}[htbp]
\centering 
\includegraphics[width=.45\textwidth]{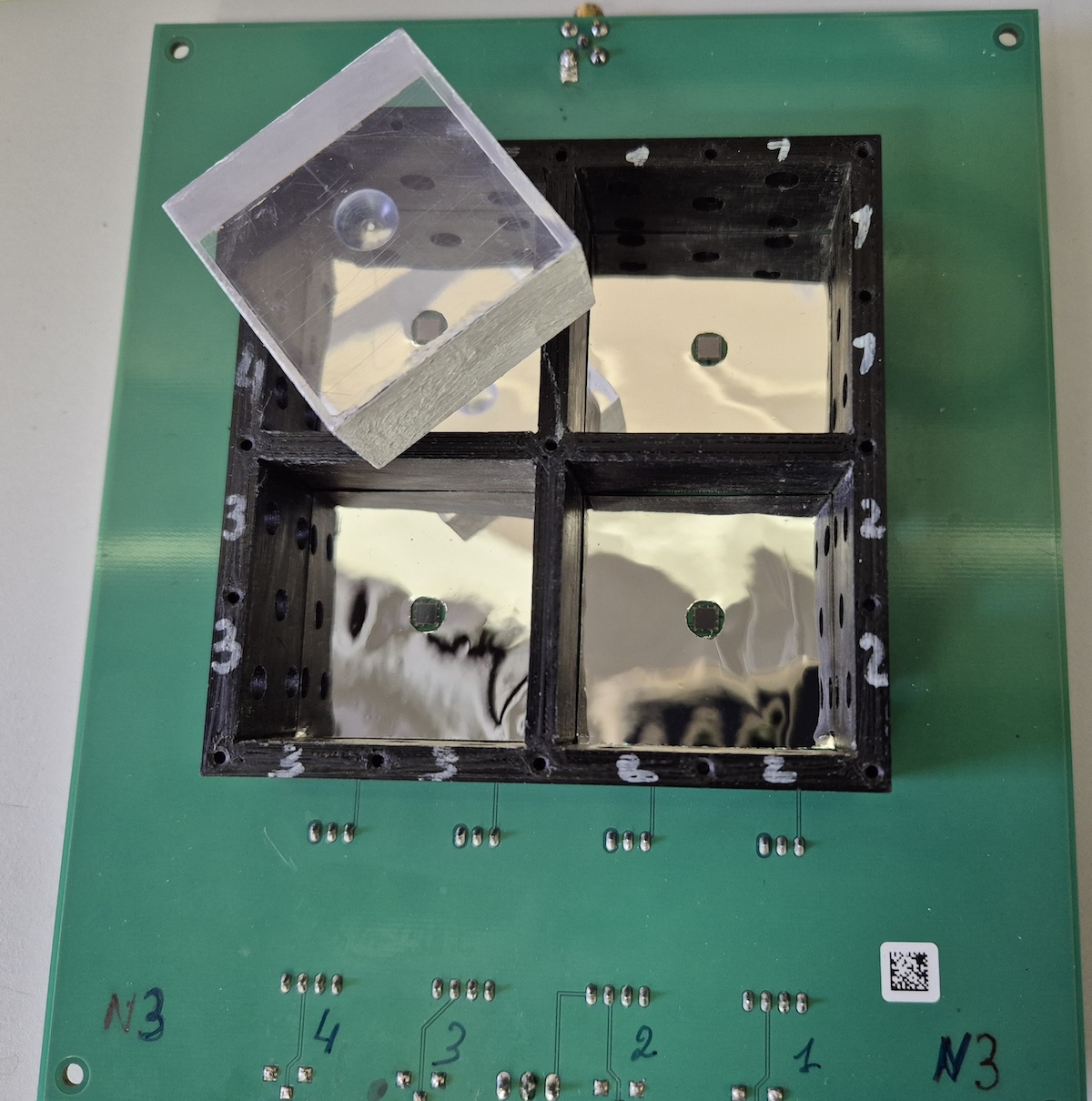}
%\qquad
%\includegraphics[width=.45\textwidth]{}
\caption{\label{fig:tile20mm} Cassette for TEST tiles and film slices with holes in front of SiPMs. Upper left the tile extracted from the cassette is shown. }
\end{figure}

The setup includes three test boards, two trigger boards and the power supply distribution board. Each test board contains four soldered SiPMs (SensL MicroFC-30035-SMT-WP with the sensitive area 3$\times$3~mm$^2$), sockets for four amplifiers and connectors for power supply and output signal readout. The black nontransparent 3D-printed cassettes to hold scintillator tiles under study are fixed on the test boards, so that each tile dimple is exactly in front of the corresponding SiPM. An example figure~\ref{fig:tile20mm} shows the cassette for TEST tiles. The cassettes are also equipped with lids to fix vertical tile positions with respect to the SiPMs. The small PCBs with LEDs are mounted on lids to provide additional illumination, which is necessary at low temperatures to get single photoelectron (SP) spectra from SiPM and extract gains for each channel.

Each trigger board contains trigger counter, an amplifier socket and connectors for power supplies and signal outputs. Both trigger counters are scintillator tiles 140$\times$140$\times$4~mm$^3$ with wavelength-shifting fibers coupled to the Hamamatsu S13360-1350PE SiPMs and are kept in the 3D-printed nontransparent cassettes to fix their position.
The horizontal area of the trigger counters perpendicular to the muon flux is larger than the total area of the test tiles on one plate.

\begin{figure}[htbp]
\centering 
\includegraphics[width=0.9\textwidth]{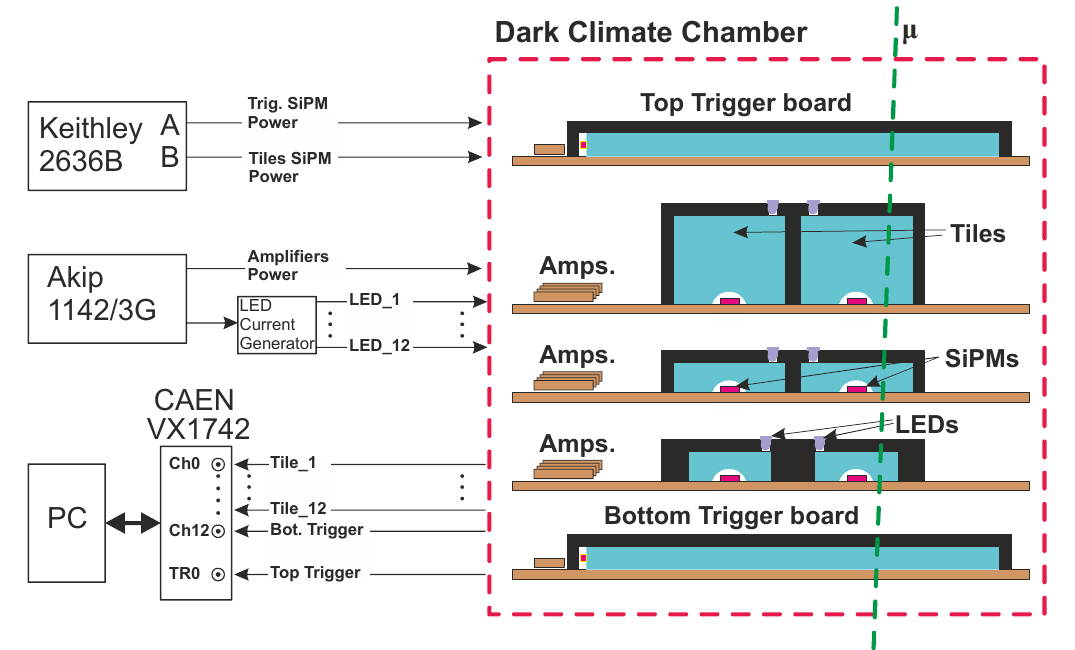}
\caption{\label{fig:rack} Schematic view of the test setup (not in scale). The vertical slice of the rack is shown. The most bottom power distribution plate is not shown.}
\end{figure}

The dedicated rack for tiles measurements is 180$\times$160$\times$210~mm$^3$ and has six horizontal guide grooves. The schematic view of experimental setup is shown on figure~\ref{fig:rack}. The power supply distribution board is in the most bottom position in the rack. The next position holds the bottom trigger counter, above which three test boards are placed on top of each other. The most upper position holds the upper trigger counter, the vertical distance between trigger counters being approximately 130~mm. 

To measure a response to muons, the whole rack is put in the dark climate chamber with temperature control precision of 0.1\textdegree C. Before each measurement, the setup has been kept during more than 10 hours at a given temperature for the temperature stabilisation to be achieved. The power supply and readout cables went outside the chamber through the dedicated technological hole. The Dual-channel Source Measure Unit Keithley-2636B is used for SiPM power supply, the same bias voltage from one output being set for 12 tile SiPMs and the same bias voltage from another output -- for 2 trigger SiPMs. The triple-channel power supply AKIP-1142/3G provides power for the amplifiers and LED current generator. The data acquisition and digitisation are performed by the 32-channel 12-bit flash-ADC CAEN~VX1742. The top trigger counter is connected to the fast trigger input of the flash-ADC, while another 13 channels including the bottom trigger counter are connected to the usual flash-ADC inputs.

\subsection{Data taking and processing}
\label{sec:processing}
A response to cosmic muons has been measured at five temperature points from $-$30\textdegree C to $+$30\textdegree C with the step of 15\textdegree C. Each data taking at a given temperature starts with illumination adjustment and calibration procedures. The illumination by LEDs is adjusted so that the "dark" current at each test channel is approximately 200--250~nA at the preselected SiPM bias voltages. The calibration is a measurement of noise spectra for all test tile channels in asynchronous mode (random trigger) at five bias voltages, $V_{\mathrm{bias}}$, around the selected operating voltage. Neither top nor bottom trigger participate in the calibration measurements. The main goal of calibration is channel-wise estimation of breakdown voltages and optical crosstalk probabilities as described in section \ref{sec:xt} below. The information about breakdown voltage is necessary to control the overvoltage, which is defined as a difference between operating and breakdown voltages of a SiPM, because the overvoltage affects the SiPM photodetection efficiency. 

During muon data taking, the readout of 14 waveforms from the flash-ADC is initiated when the top trigger amplitude exceeds the preset threshold defining the event timestamp. All the 14 waveforms (sampled with 1~GHz frequency) are recorded for further analysis if the bottom trigger amplitude also exceeds its threshold within 10 ns from the top trigger timestamp. For the given setup layout, each triggered cosmic muon crosses zero, one, two or three test tiles resulting in corresponding combination of muon and noise signals in the test channels. Thus, the noise spectra for gain estimates are collected simultaneously with the signal spectra from cosmic muons. About 12000 triggered events are typically recorded during 5 hours of data taking providing approximately 1400 muon signals per TEST tile and about 1000 muon signals per AHCAL tile because tile areas perpendicular to muon flux are different.

The data processing starts from the waveform analysis. For each test channel, the signal from muon event is calculated as an integral over the waveform from the top trigger timestamp in the fixed gate of 235~ns  and the integral is then divided to the gate length and presented in units of normalised ADC counts (nADC). The integration gate was defined based on the pulse length at the lowest studied temperature point, $-$30\textdegree C. The typical averaged pulse shape in one channel is shown in figure~\ref{fig:pulse_sps} (left). A muon signal in the tiles under study exceeds 20 photoelectrons and can be easily separated from the noise signal by a simple channel-wise threshold cut. Then, the noise and muon events are analysed separately as described below.

\subsubsection{Gain extraction}
\label{sec:gain}
\begin{figure}[htbp]
\centering 
\includegraphics[width=.45\textwidth]{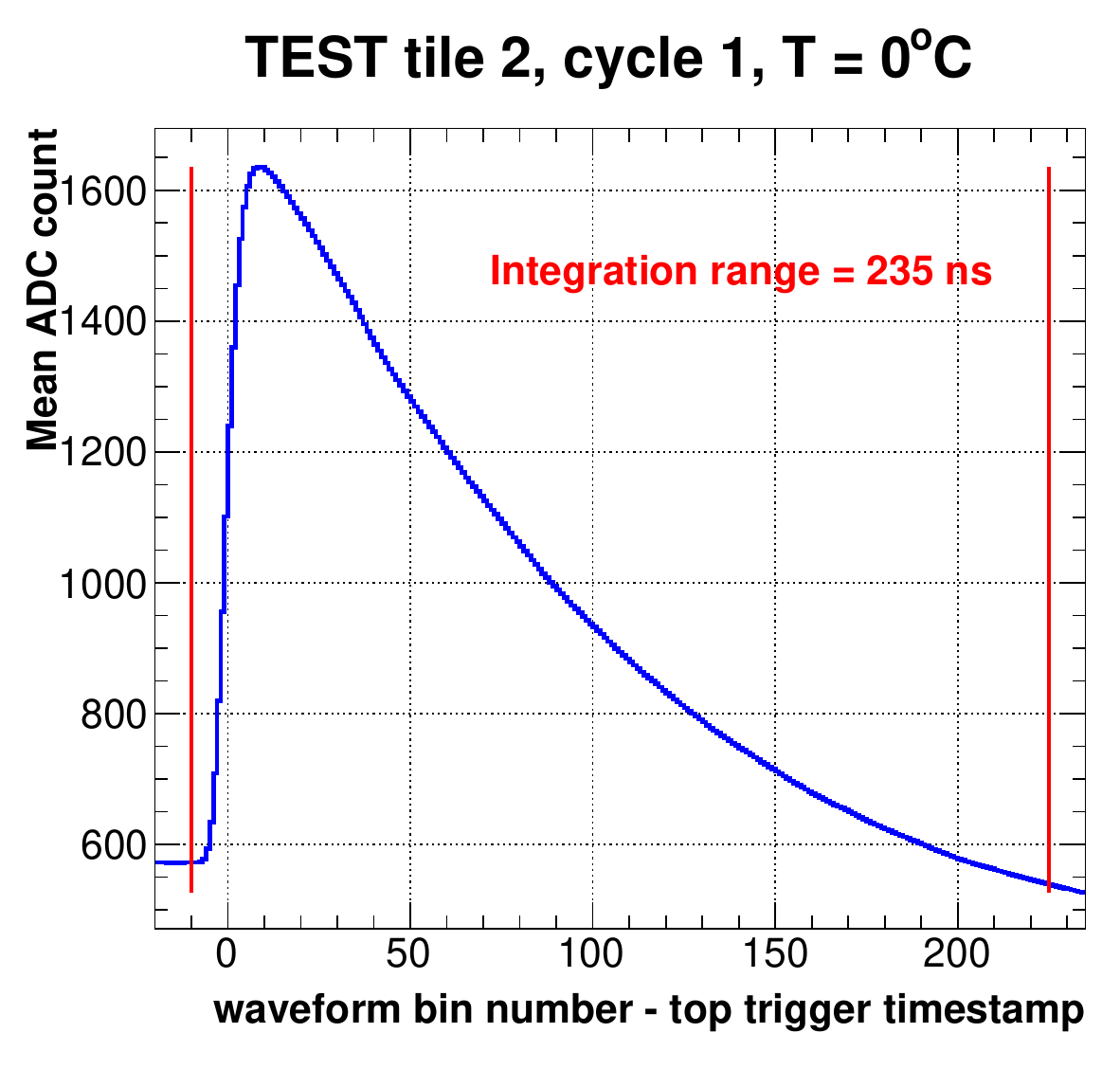}
\qquad
\includegraphics[width=.45\textwidth]{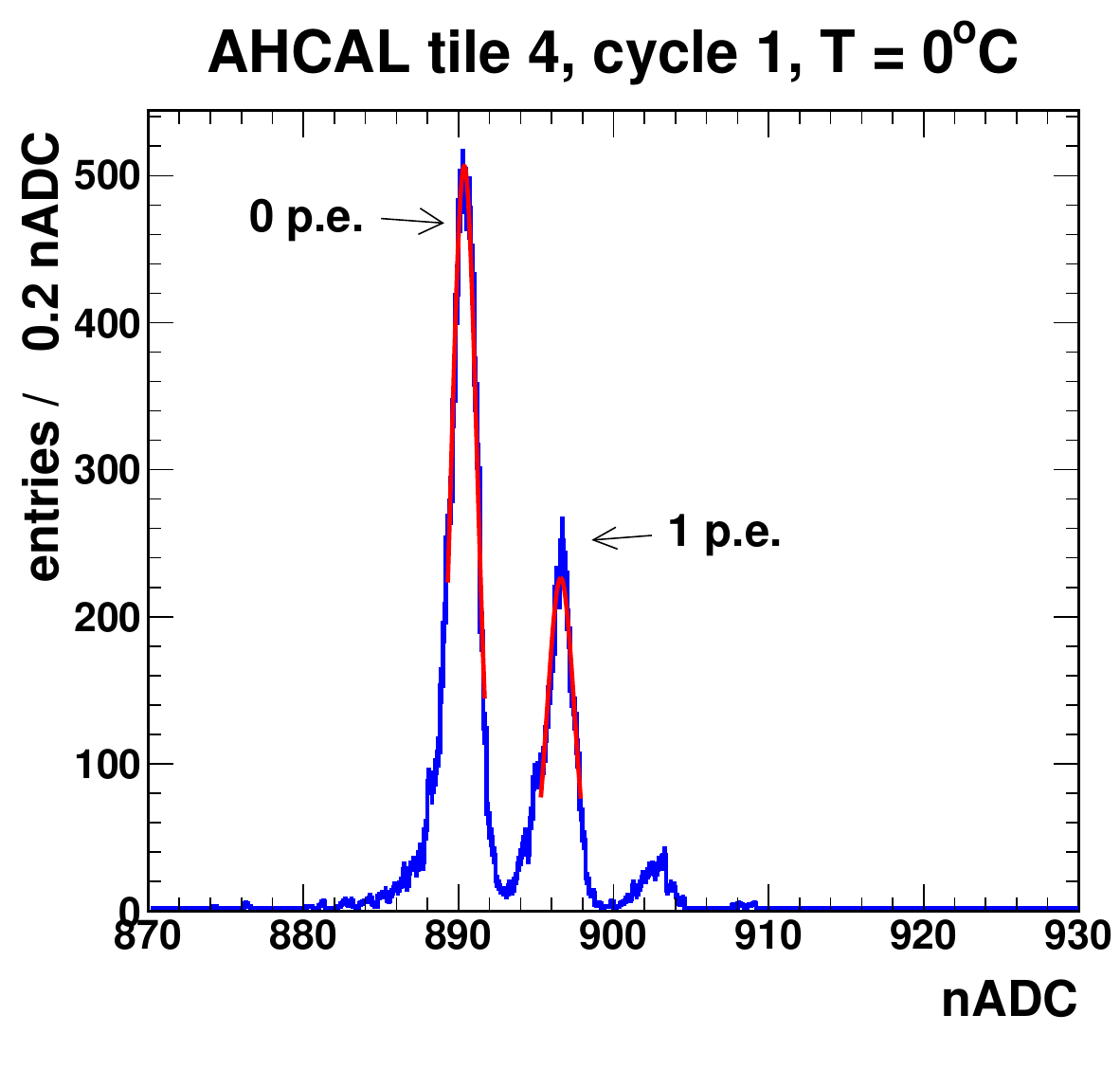}
\caption{\label{fig:pulse_sps} Left: average pulse shape of muon signal (blue); red lines show integration range. Right: typical noise spectrum enriched by single photoelectron pulses (blue); red curves correspond to Gaussian fits.}
\end{figure}

The channel-wise estimation of gain helps to minimise the influence of SiPM behaviour on the light yield measurements. To extract gain, a well resolved single photoelectron spectrum is necessary. The dedicated algorithm has been developed for the search of SP pulses in the 1-$\mu$s-long recorded waveform for events, where no muon signal is detected. The typical SP-enriched spectrum is shown in figure~\ref{fig:pulse_sps} (right). Both pedestal and single photoelectron peaks of the enriched spectra were fit independently with Gaussian functions resulting in the values $SP0$ and $SP1$ of the corresponding means. Then the gain is determined as $G = SP1 - SP0$ in units of nADC/p.e. The systematic uncertainties related to fit ranges and binning do not exceed 0.3\%. The overall relative uncertainty of gain estimates was 1--2\%. 

The same procedure of waveform analysis and gain extraction is applied to both calibration and cosmic data. From calibration measurements, the SiPM breakdown voltages have been estimated using the standard method of extrapolation  of the linear dependence of gain on bias voltage to the point $G = 0$~\cite{ref:klanner}. The uncertainty of bias voltage settings (50~mV) and gain uncertainties are taken into account in these linear fits. Then the overvoltages, $OV_{ij}$, for $i$-th channel at $j$-th temperature are calculated. Typical uncertainties of overvoltage estimates are about 1--2\%    

\subsubsection{Optical crosstalk estimation}
\label{sec:xt}
The optical crosstalk between SiPM cells is defined as an additional cell firings initiated by secondary photons from the neighbour cell fired~\cite{ref:klanner}. The optical crosstalk contributes to the measured light yield values. 
Let us assume that the probability $P(n)$ of firing $n$ cells follows the Poisson distribution with the mean $\mu$. Then the following relations can be written:
\begin{equation}
\label{eq:p0}
P(0) = e^{-\mu} = \alpha,  \quad P(1) = -\alpha \ln(\alpha) \cdot (1 - c), \quad P(n>1) = 1 - P(0) - P(1),
\end{equation}

\noindent where $c$ is the optical crosstalk probability. The probabilities $P(0)$, $P(1)$ and $P(n>1)$ can be estimated from noise spectrum with total number of events, $N_{\mathrm{total}}$, as number of events fractions in the corresponding ranges:

\begin{equation}
P(0) \approx \frac{N \in [L_{min}; L_0]}{N_{\mathrm{total}}}, \quad
P(1) \approx \frac{N \in [L_0; L_1]}{N_{\mathrm{total}}}, \quad
P(n > 1) \approx \frac{N \in [L_{1}; L_{max}]}{N_{\mathrm{total}}}, \\
\end{equation}
\begin{equation*}
L_{0} = SP0 + 0.5\cdot G, \quad L_{1} = SP1 + 0.5\cdot G,
\end{equation*}

\noindent where  $[L_{min}, L_{max}]$ is the full interval of noise spectrum and values $SP0$, $SP1$ and $G$ are defined in the previous section. From eq.~\ref{eq:p0} the crosstalk estimate can be derived as follows: 

\begin{equation}
\label{eq:xt}
c = 1 + \frac{1 - \alpha}{(1 + q) \cdot \alpha \cdot \ln(\alpha)}, \quad q = \frac{P(n>1)}{P(1)}.
\end{equation}

While the values of $\alpha$ should be calculated from raw spectra, a more precise estimate of $q$ can be obtained from the SP-enriched spectra assuming the same efficiency of finding single or multiple photoelectron signal. The optical crosstalk probabilities for the SiPMs of test channels were estimated to be $\sim$10\% at 2.5~V overvoltage and $\sim$20\% at 5~V overvoltage with quite weak temperature dependence. The relative uncertainty of crosstalk estimates is about 10\%.

\subsubsection{Relative efficiency of photon detection}
The photon detection efficiency (PDE) depends on the overvoltage applied to a SiPM. Even for the SiPMs of the same type, there is a spread of breakdown voltages declared by the manufacturer, which results in difference of overvoltages for the same bias voltage applied. This difference amounts to 450~mV for the SiPMs in the described setup. In this study, a temperature dependence of response to minimum ionising particle is estimated per channel without any channel-to-channel comparison. Nevertheless, there might be a spread of temperature dependence slopes, which leads to different overvoltages at different temperatures for a particular SiPM if the increment of bias voltage with temperature is fixed for all channels (SiPMs). Such a difference is observed in the current setup and is taken into account in the studies of light yield temperature dependence.

The coarse version of the PDE dependence on overvoltage is usually provided by manufacturers. This dependence was studied in more details in ref.~\cite{ref:gundacker} for the SensL SiPMs similar to those used in the current experiment. The relative correction factor, $r_{ij}$, for $i$-th channel at $j$-the temperature point is calculated from the parametrisation of the $PDE(OV)$ curve from ref.~\cite{ref:gundacker} as follows: 

\begin{equation}
\label{eq:pde}
r_{ij} = \frac{PDE(OV_{i\mathrm{ref}})}{PDE(OV_{ij})},
\end{equation}

\noindent where $OV_{i\mathrm{ref}}$ is the overvoltage at the reference temperature that is chosen to be $T_{\mathrm{ref}} = -$30\textdegree C. The uncertainty of this factor gives the noticeable contribution to overall uncertainty at the level of 1.5--2.5\%
s
\subsubsection{Fit of muon signal distributions and light yield calculation}
A typical distribution of signal from cosmic muons at 0\textdegree C is shown in figure~\ref{fig:mip} for 20-mm-thick TEST tile (left) and for 3-mm-thick AHCAL tile (right). The distribution for thicker tile is noticeably wider as expected. The fit function shown is a convolution of Landau and Gaussian distributions. The convolution function has four free parameters: the most probable value (LMP) and width (Lwidth) of Landau distribution, the Gaussian width (Gsigma) and normalisation factor (Area), which are shown in the legends of figure~\ref{fig:mip} for the particular fits. The response to minimum ionising particle is defined as a maximum of the convolution function, $M_{ij}$, for $i$-th channel at temperature $T_{j}$. The systematic uncertainties related to the fit range and bin width variations have been investigated and were found to be $<$0.3\%. The overall relative uncertainty of response to muons $M_{ij}$ varies from 0.5\% to 1\%. The light yield, $LY_{ij}$, of channel $i$ at temperature $T_{j}$ is calculated as follows:

\begin{figure}[htbp]
\centering 
\includegraphics[width=.45\textwidth]{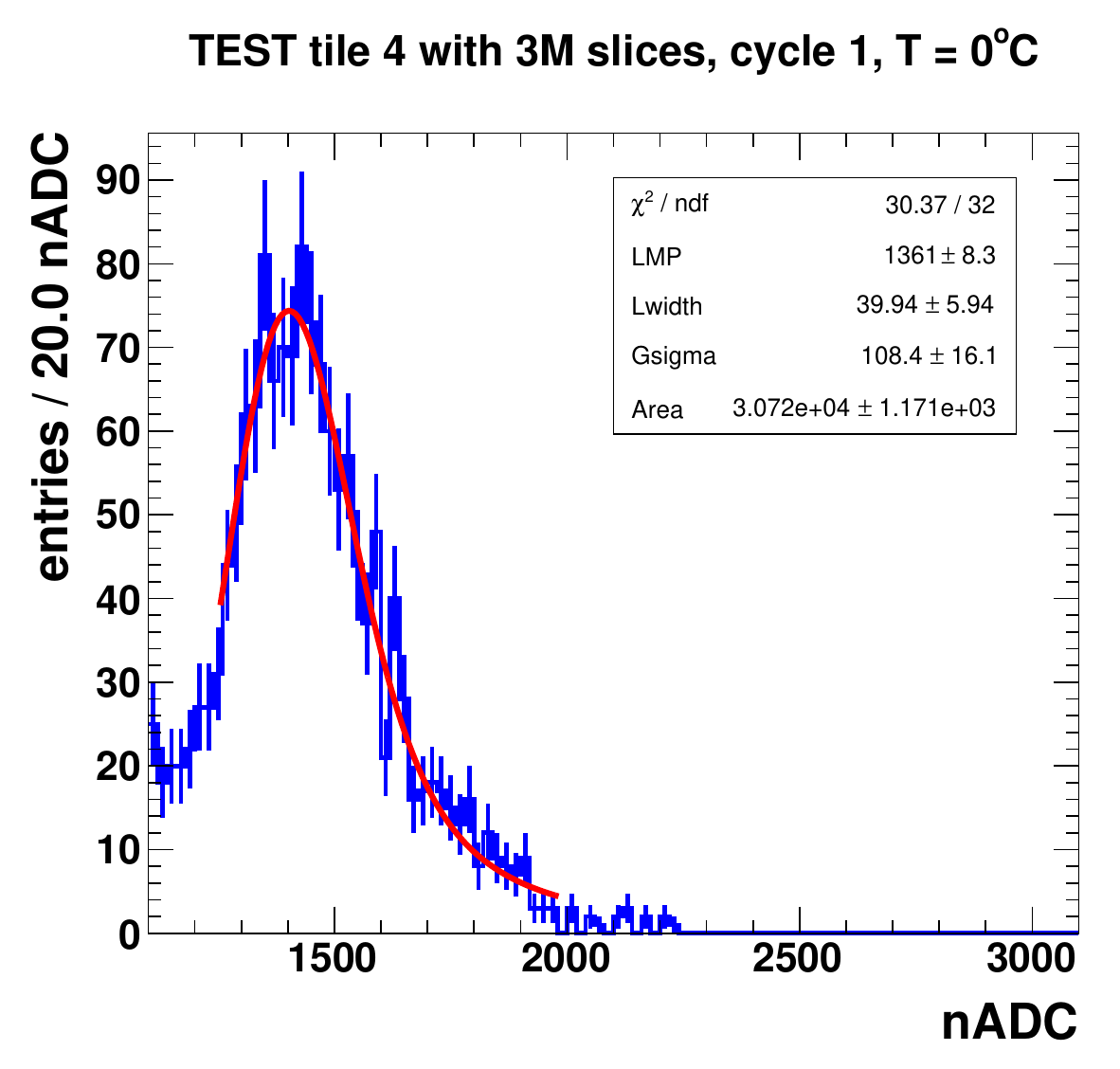}
\qquad
\includegraphics[width=.45\textwidth]{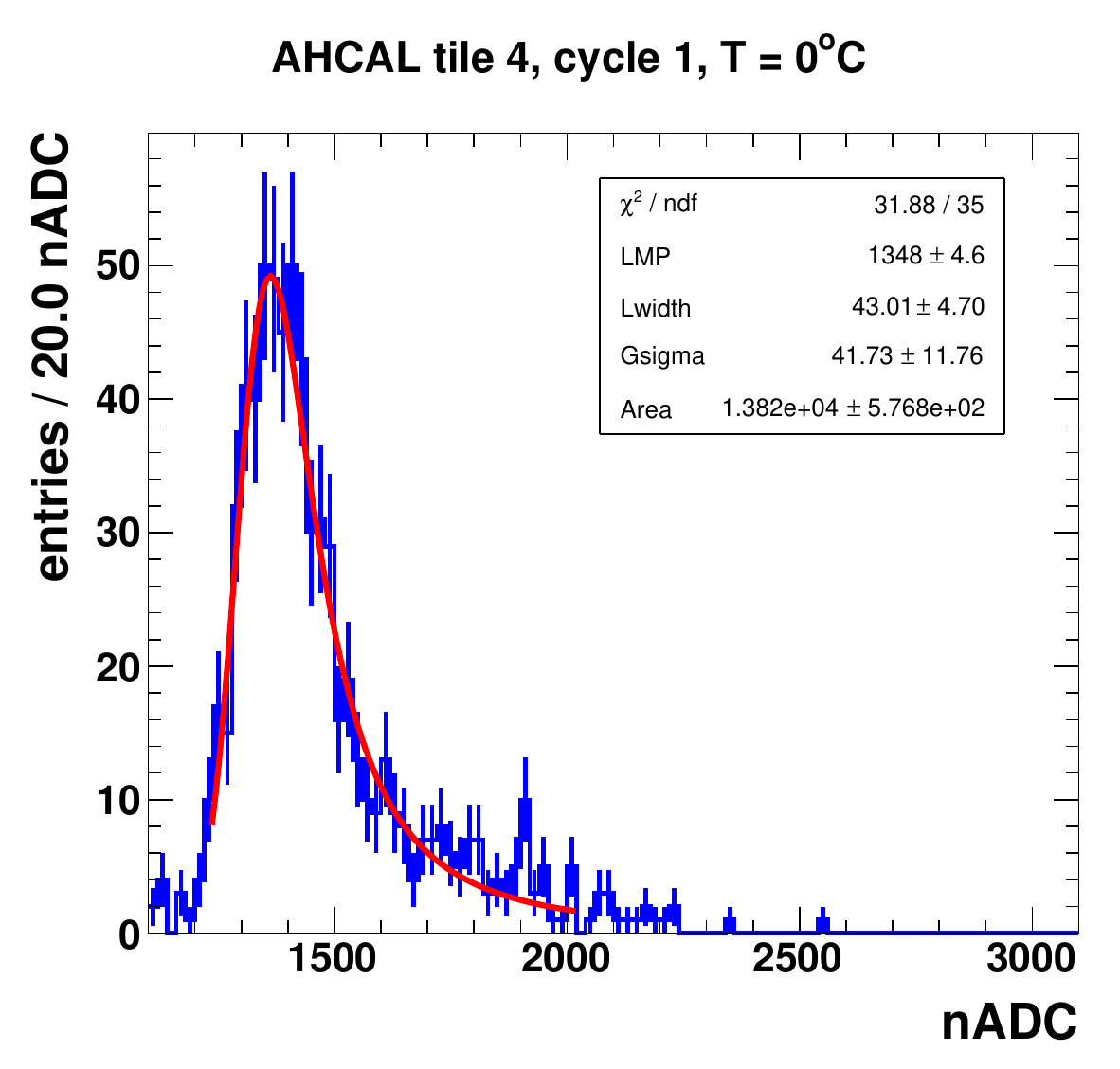}
\caption{\label{fig:mip} Typical response to cosmic muons (blue) for TEST tile (left) and AHCAL tile (right); red curves correspond to the fit with Landau-Gaussian convolution function (see text for details).}
\end{figure}

\begin{equation}
\label{eq:ly}
LY_{ij} = \frac{ (M_{ij} - P0_{ij}) \cdot r_{ij} }{ G_{ij} \cdot (1 + c_{ij}) }.
\end{equation}

The correction factors applied are aimed at mitigation of any temperature dependences of photodetector and electronics properties, such as gain, crosstalk contribution and photon detection efficiency estimated using eq.~\ref{eq:xt} and \ref{eq:pde}. The uncertainty of single light yield measurement for single channel varies from 2\% to 4\%.

\section{Results}
\label{sec:res}
\begin{figure}[htbp]
\centering 
\includegraphics[width=.48\textwidth]{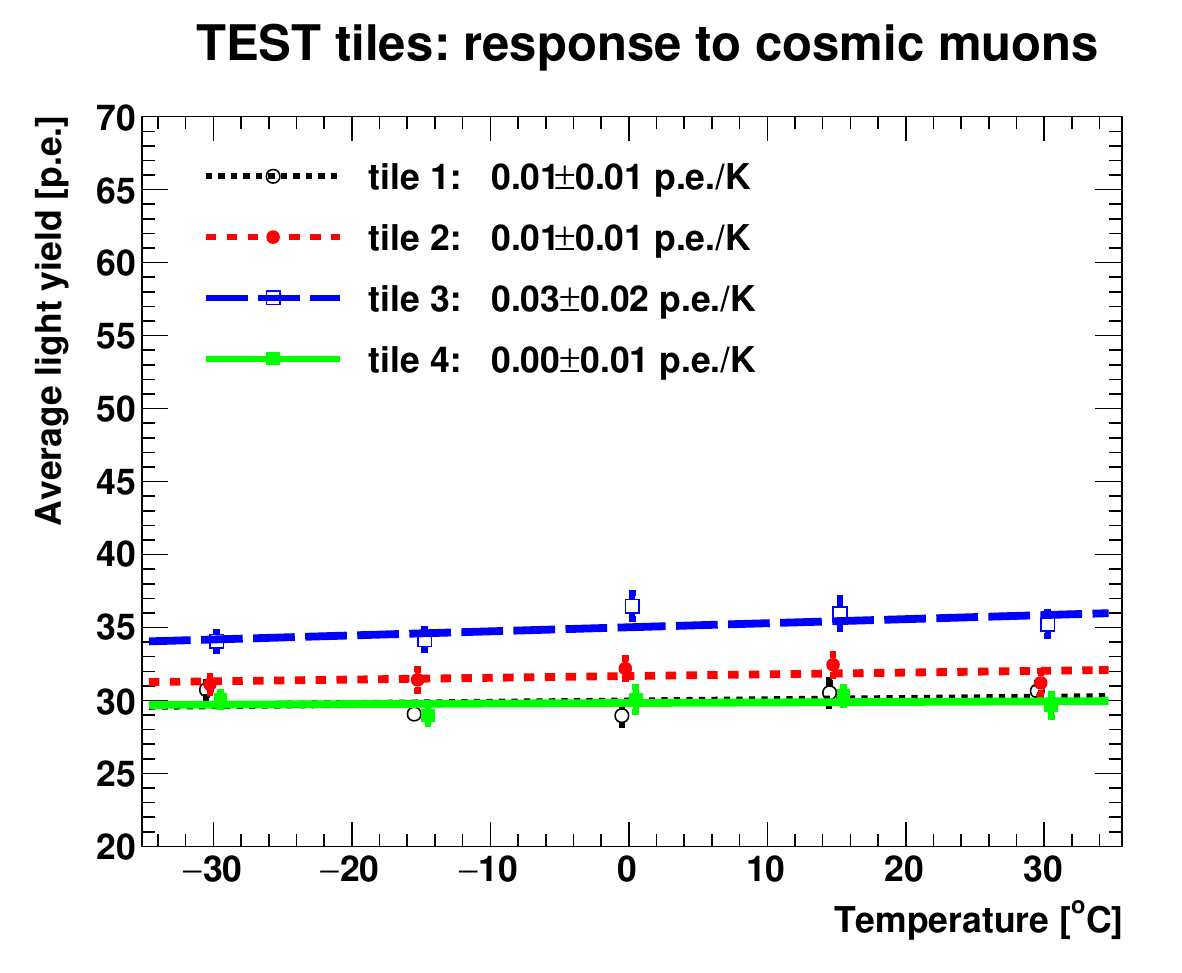}
\includegraphics[width=.48\textwidth]{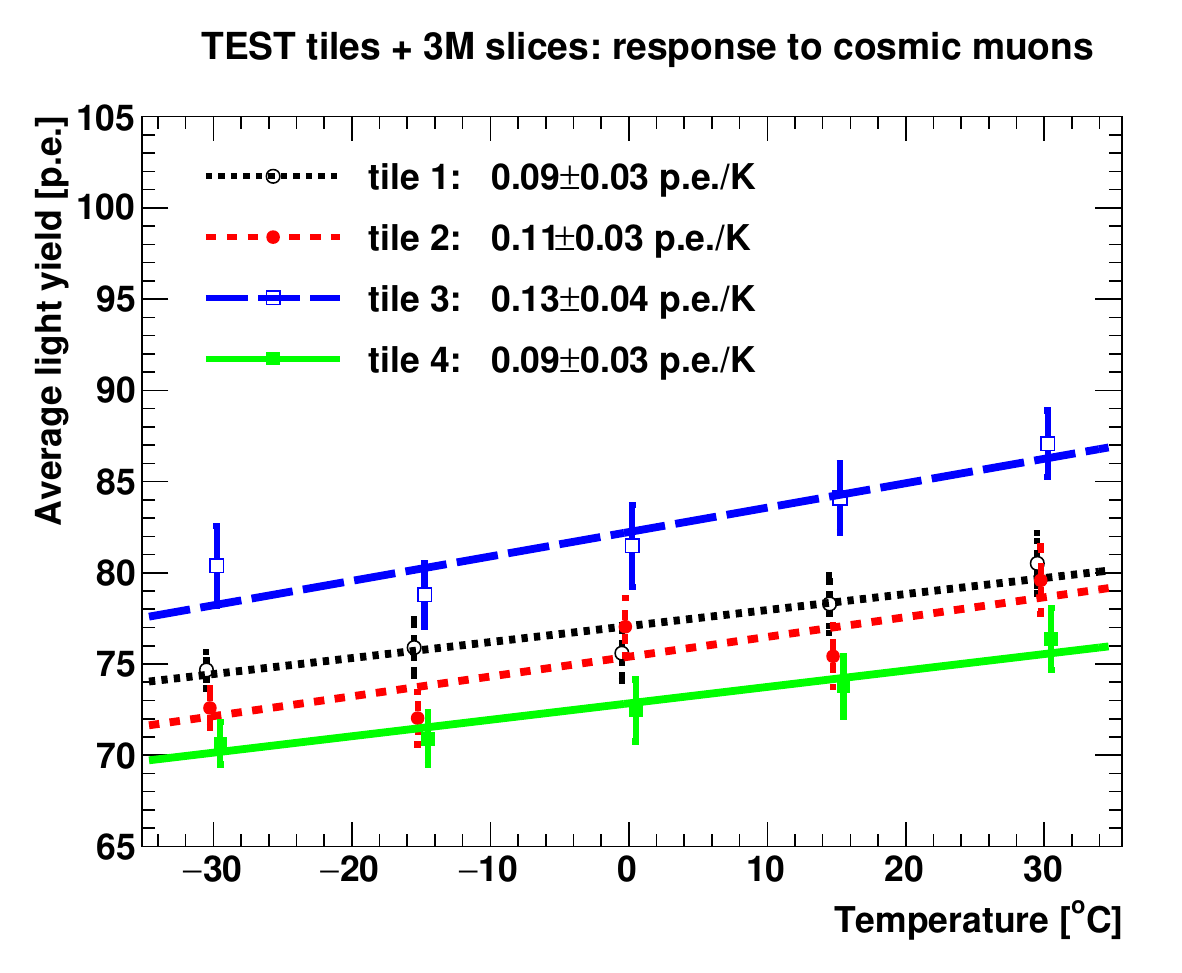}
\caption{\label{fig:testtiles} Response to cosmic muons versus temperature for TEST tiles without (left) and with 3M slices of reflective film (right). Error bars show error of mean calculated over two heating-cooling cycles. Lines correspond to linear fits with slopes shown in the legend. Points are slightly shifted along $x$ axis for better visibility.}
\end{figure}

For each composition of tile types, the measurements of response are repeated for several heating-cooling cycles in the temperature range from $-$30\textdegree C to $+$30\textdegree C with the same bias voltage settings at a given temperature. The response to cosmic muons was calculated using eq.~\ref{eq:ly} for each channel and temperature point and then averaged over cycles. The measurements of four bare TEST tiles have been performed at overvoltages about 5~V, the full heating-cooling cycle has been repeated twice.  As can be seen from figure~\ref{fig:testtiles} (left), no temperature dependence of response has been observed for bare tiles. The linear fits give slopes consistent with zero and constant fits are consistent with data. The difference between tiles in absolute response can be explained by manual production of tiles including manual dimple polishing and side coating by diffuse reflector. 

At the next step, the slices of 3M ESR film have been placed below and above the TEST tiles and their response has been measured during two full heating-cooling cycles. For these measurements, the tiles have being placed at the same positions in the cassette, i.e. in front of the same SiPMs, as in the previous measurements without film slices and the overvoltage has been reduced to 2.5~V to avoid ADC overflow. The increase of response with increasing temperature can be observed in figure~\ref{fig:testtiles}. The slope of response temperature dependence amounts to $\sim$0.1~p.e./K on average and constant fits are not consistent with data.

\begin{figure}[htbp]
\centering 
\includegraphics[width=.48\textwidth]{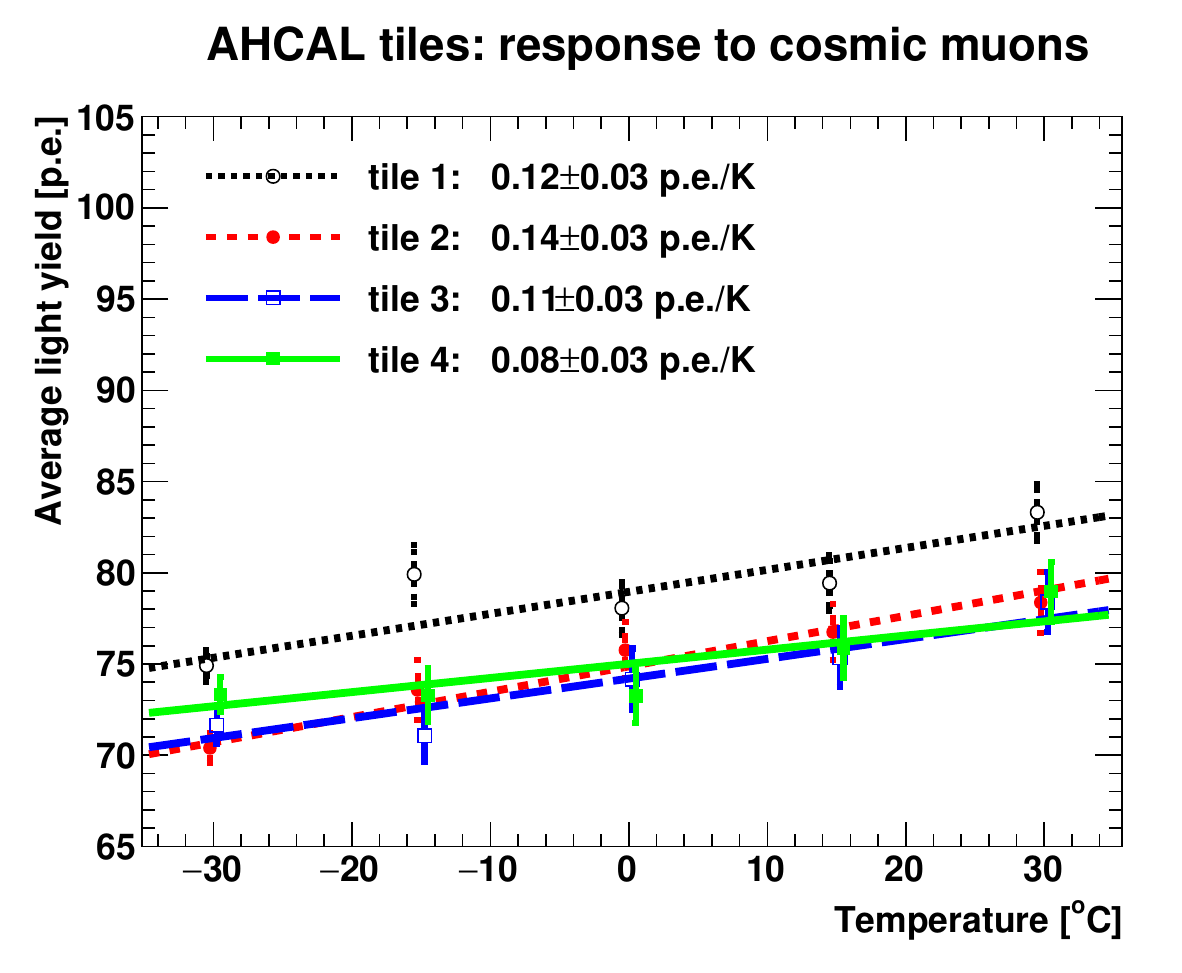}
\includegraphics[width=.48\textwidth]{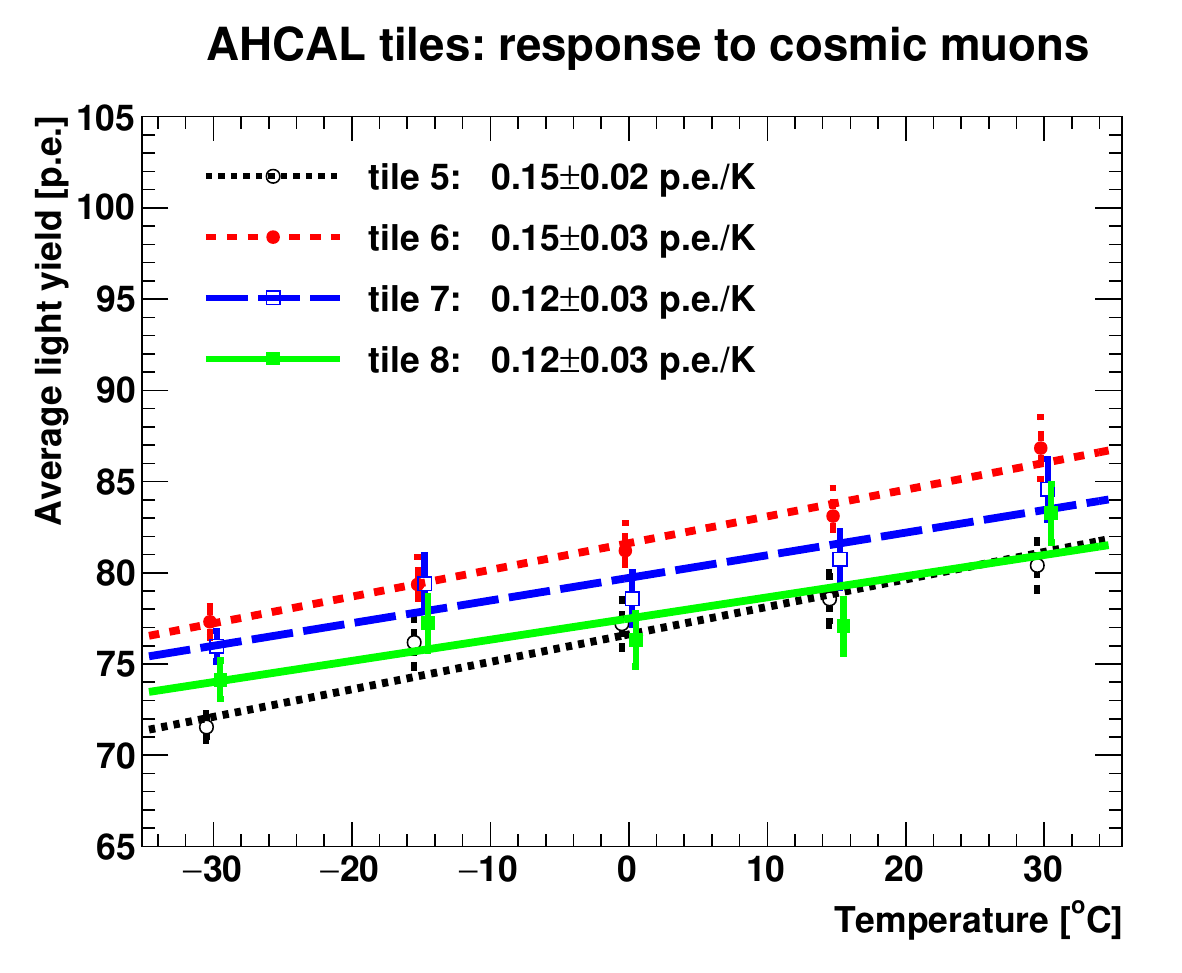}
\caption{\label{fig:calice} Response of AHCAL tiles to cosmic muons versus temperature: tiles 1--4 (left) and tiles 5--8 (right). Error bars show error of mean calculated over two heating-cooling cycles. Lines correspond to linear fits with slopes shown in the legend. Points are slightly shifted along $x$ axis for better visibility.}
\end{figure}

Eight AHCAL tiles have been also measured at 2.5~V overvoltage using two full heating-cooling cycles. The AHCAL tiles have smaller area perpendicular to muon flux resulting in about 30\% smaller number of collected muon events than for TEST tiles. They show similar temperature dependence of response to muons as TEST tiles covered by 3M slices as can be seen in figure~\ref{fig:calice}, where results are presented for AHCAL tiles randomly divided into two groups for better graphic representation. The spread of response between AHCAL tiles at each temperature does not exceed 6\%.

\section{Conclusion}
\label{sec:conclusion}

The temperature dependence of light yield of plastic scintillator tiles has been studied for different options of tile combinations with reflective film by measuring the response to cosmic muons. The scintillation light from tiles was detected by silicon photomultipliers. The impact of SiPM and other electronics is mitigated by taking into account channel-wise estimations of gain, crosstalk contribution and relative photodetection efficiency at each temperature point. For all options under study, the response to muons has been measured at five temperature points in the range from $-$30\textdegree C to $+$30\textdegree C.  

For all tile combinations with reflective film, either slices or wrapping usage, the increase of response to minimum ionising particles with temperature has been observed in the temperature range studied. The effect is measured to be about 1~p.e. (or 1.3\% of response) per 10~K and is very similar for three investigated tile-film configurations: few specimens of mass produced and automatically wrapped square tiles for the CALICE AHCAL technological prototype, for the manually produced 20-mm thick square tiles covered by film slices and for previously measured manually wrapped specimens of trapezoidal HGCAL tiles~\cite{ref:bull2025}. The overall observed increase of light yield is $\sim$10\% for temperature increment by 60\textdegree C.

The 20-mm thick scintillator tiles have been also measured without reflective film to understand the contributions from constituents to the effect observed. It was found that the bare TEST tiles with just diffuse reflector coating on the perimeter side show no temperature dependence of light yield. The behaviour changes when reflective film slices are added  -- the combination of the same tiles with reflective film slices results in appearance of light yield increase with temperature. The effect can be explained by reflective film properties that might degrade with cooling but this is not supported by information from manufacturers. The effect observed is most likely the result of complicated interrelation between the particular materials: plastic scintillator and reflective film. Therefore, tests of such assemblies at the expected operating conditions will help to increase the reliability of performance studies.

\acknowledgments
We are grateful to Vladimir Rusinov for the production of TEST tiles and to the CALICE colleagues for providing the AHCAL wrapped tiles. The development and testing of electronic components for this study was supported by the Ministry of Science and Higher Education of the Russian Federation, Project "New Phenomena in Particle Physics and the Early Universe" FSWU-2023-0073.

\bibliographystyle{JHEP}
\bibliography{chadeeva_arxiv.bib}

\end{document}